\documentclass[iop,floatfix,numberedappendix,twocolappendix]{emulateapj}

\usepackage[backref,breaklinks,colorlinks,urlcolor=blue,citecolor=blue,linkcolor=blue]{hyperref}
\usepackage[all]{hypcap}

\usepackage{graphicx}
\usepackage{enumerate}
\usepackage{amsmath,amssymb}
\usepackage{bm}
\usepackage{color}
\usepackage[utf8]{inputenc}

\usepackage[normalem]{ulem}

\begin{document}

\title{A Disk-based Dynamical Mass Estimate for the Young Binary AK Sco}
\author{I.~Czekala\altaffilmark{1}, S.~M.~Andrews\altaffilmark{1}, E.~L.~N.~Jensen\altaffilmark{2}, K.~G.~Stassun\altaffilmark{3,4}, G.~Torres\altaffilmark{1}, \& D.~J.~Wilner\altaffilmark{1}}
\altaffiltext{1}{Harvard-Smithsonian Center for Astrophysics, 
			 60 Garden Street, Cambridge, MA 02138; \email{iczekala@cfa.harvard.edu}}
\altaffiltext{2}{Department of Physics and Astronomy, Swarthmore College, 500 College Avenue, Swarthmore, PA 19081}
\altaffiltext{3}{Department of Physics and Astronomy, Vanderbilt University, Nashville, TN 37235}
\altaffiltext{4}{Department of Physics, Fisk University, Nashville, TN 37208}

\def\teff{$T_{\rm eff}$}
\def\logg{$\log g$}
\newcommand{\radmc}{\texttt{RADMC-3D}}
\newcommand{\kms}{ \textrm{km s}^{-1} }
\newcommand{\todo}[1]{ \textcolor{red}{#1}}
\newcommand{\vt}{ {\bm \theta}}
\newcommand{\msun}{M$_\odot$}

\begin{abstract}
We present spatially and spectrally resolved Atacama Large Millimeter/submillimeter Array (ALMA) observations of gas and dust in the disk orbiting the pre-main sequence binary AK Sco. By forward-modeling the disk velocity field traced by CO $J$=2$-$1 line emission, we infer the mass of the central binary, $M_\ast = 2.49 \pm 0.10$\,$M_\odot$, a new dynamical measurement that is independent of stellar evolutionary models.  Assuming the disk and binary are co-planar within $\sim$2\degr, this disk-based binary mass measurement is in excellent agreement with constraints from radial velocity monitoring of the combined stellar spectra.  These ALMA results are also compared with the standard approach of estimating masses from the location of the binary in the Hertzsprung-Russell diagram, using several common pre-main sequence model grids.  These models predict stellar masses that are marginally consistent with our dynamical measurement (at $\sim$2\,$\sigma$), but are systematically high (by $\sim$10\%).  These same models consistently predict an age of $18\pm1$\,Myr for AK Sco, in line with its membership in the Upper Centaurus-Lupus association but surprisingly old for it to still host a gas-rich disk.  As ALMA accumulates comparable data for large samples of pre-main sequence stars, the methodology employed here to extract a dynamical mass from the disk rotation curve should prove extraordinarily useful for efforts to characterize the fundamental parameters of early stellar evolution.   
\end{abstract}
\keywords{ protoplanetary disks -- stars: fundamental parameters -- stars: pre-main sequence -- stars: individual (AK Sco)}

\section{Introduction \label{sec:intro}}

Precise measurements of the physical properties of pre-main sequence (pre-MS) stars are fundamental to testing the theoretical predictions of stellar evolution models.  Such models are in turn the basis for deriving a variety of interesting quantities in star-formation research, including the masses and ages of individual pre-MS stars from secondary properties (such as effective temperature and luminosity), the initial mass function (IMF) of star-forming regions, and the timescale for circumstellar disk evolution and planet formation, among many others.  To successfully test and help refine theoretical models of pre-MS evolution, we require a sample of ``benchmark" systems where the fundamental properties are known.  The key constraint for such systems is a direct (i.e., dynamical) measurement of the stellar mass.

Unfortunately, few such benchmarks exist, particularly at the low end of the mass spectrum ($M_\star < 2$ \msun) where the models are in greatest need of calibration \citep[see, e.g.,][]{Hillenbrand2004,Mathieu2007,Stassun2004,Stassun2006,Stassun2007,Stassun2008,Gennaro2012}.  Eclipsing binaries (EBs) have long served as important empirical touchstones for testing stellar models.  However, the recent review of pre-MS benchmarks by \citet{Stassun2014} identified only 21 low-mass EBs that have sufficiently precise measurements of their physical parameters to be suitable for testing evolutionary models.  The same review also presented new evidence that many of the benchmark pre-MS EBs may have their temperatures and/or radii altered by the influence of tertiary companions which, while representing interesting physics in their own right, render them less suitable to direct tests of the evolutionary models that do not include such effects \citep[see also][]{Gomez2012}. 

Therefore, additional pre-MS stellar mass benchmarks are crucial.  Ideally, these would include single stars, or binaries without the potentially complicating effects of a tertiary component.  In particular, single or binary pre-MS stars with circumstellar or circumbinary disks offer the opportunity to dynamically measure the stellar masses using the Keplerian orbital motion (i.e., rotation curve) of the gas disk
\citep[e.g.][]{simon00,schaefer09,rosenfeld12}.

To be sure, there are additional complications with the analysis of such systems, including accretion effects, isolating the individual component masses in the case of a close binary (since the dynamical mass is then the total binary mass), and a systematic (linear) dependence on the (still uncertain) distance.  However, such systems are much more common than the intrinsically rare EBs, and therefore represent an important opportunity to significantly expand the sample of benchmark pre-MS stars.  This is particularly compelling now, as ALMA has started to provide unprecedented sensitivity to molecular line emission from disks, and the upcoming {\it GAIA} mission is poised to give accurate distances to these systems.  Just as important as the ongoing determination of empirical masses for more of these pre-MS systems is the development of robust procedures for their careful analysis, including sophisticated molecular line modeling approaches and the incorporation of state-of-the-art statistical methods for comparing the stellar properties to current theoretical pre-MS evolution models.

\begin{figure*}[t!]
  \includegraphics[width=\textwidth]{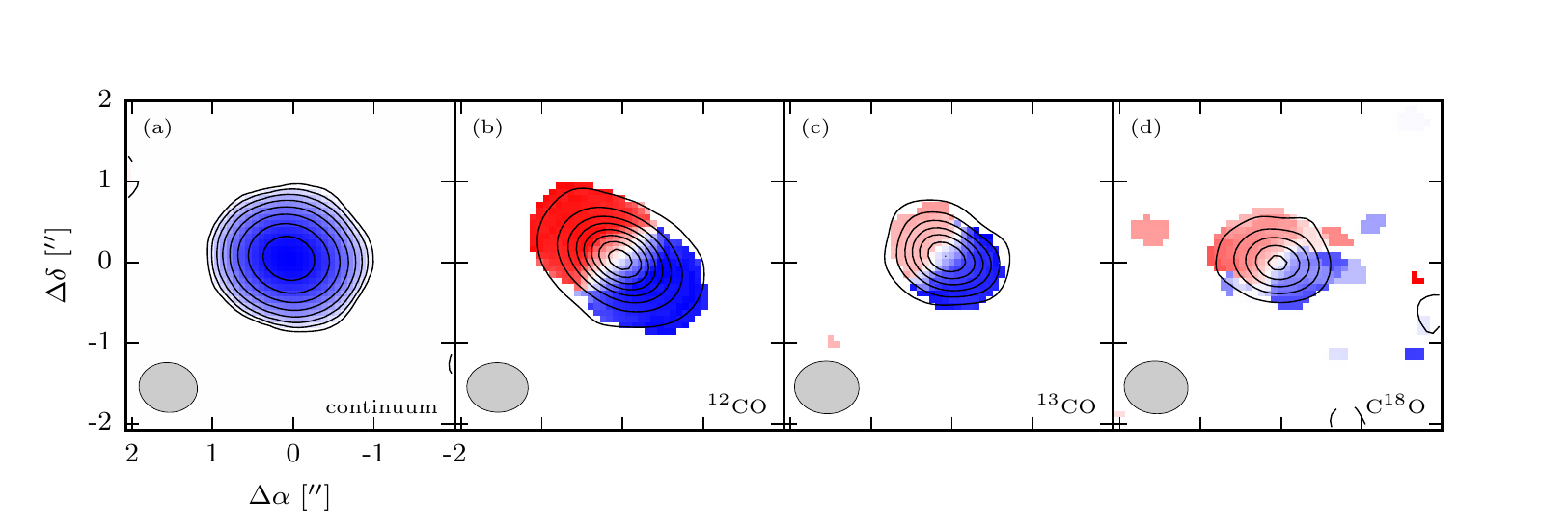}
  \figcaption{(a) Synthesized image of the 1.3\,mm dust continuum emission.  Contours start at 3\,$\sigma$ and increase by factors of two.  (b) The 0th moment map (velocity-integrated emission; contours) overlaid on the 1st moment map (intensity-weighted velocities; color scale) for the CO $J$=2$-$1 emission.  Contour levels start at 3\,$\sigma$ and increase in 10\,$\sigma$ intervals.  (c) and (d) The same as (b) for the $^{13}$CO and C$^{18}$O $J$=2$-$1 emission, respectively.  Contours are spaced at 3\,$\sigma$ intervals.  The synthesized beam is shown in the lower left of each panel.
  \label{fig:mom}}
\end{figure*}

AK Sco represents a good case study for this purpose. AK Sco is a bright ($V \approx 8.9$), pre-MS double-lined spectroscopic binary \citep{andersen89} associated with the nearby Upper Centaurus-Lupus star-forming region \citep[][$d \sim 140$ pc]{pecaut12}.  It is actively accreting and has a massive circumbinary dust disk \citep{alencar03}.   An orbital solution from long-term radial velocity monitoring of optical spectra identifies a short period ($\sim$13.6 days), eccentric ($e = 0.47$), nearly equal-mass pair of F5 stars \citep{andersen89,alencar03}.  \citet{alencar03} presented the most comprehensive analysis of the system to date, including an attempt to determine the stellar masses by constraining the (unknown) orbital inclination using geometric arguments (the stars are not known to eclipse) together with a simple model of the spectral energy distribution (SED).  Recently, \citet{anthonioz15} resolved the inner edge of the AK Sco circumbinary disk with the VLT interferometer.  However, until now there has not been a dynamical measurement of the disk rotation curve with which to directly measure the binary mass.

Here we report the direct determination of the AK Sco binary mass through the dynamical measurement of its circumbinary disk rotation profile using new data from ALMA. Section \ref{sec:data} presents the ALMA observations and their calibration.  Section \ref{sec:method} covers the methodology for modeling the disk gas rotation profile, and thereby the binary mass.
Section \ref{sec:disc} discusses these results together with an analysis of other basic stellar properties in the context of pre-MS evolutionary model predictions.  

\section{Observations and Data Reduction}\label{sec:data}

AK Sco was observed using 32 ALMA antennas and the Band 6 receivers on 2014 April 10 as part of program 2012.1.00496.S (PI Andrews), with baseline lengths of 15--350\,m.  The ALMA correlator was configured to process four dual-polarization spectral windows.  Two windows, covering the CO (230.538\,GHz) and $^{13}$CO (220.399\,GHz) $J$=2$-$1 transitions, spanned a 234\,MHz bandwidth with 61\,kHz (80\,m s$^{-1}$) channels.  A third window sampled C$^{18}$O $J$=2$-$1 (219.560\,GHz) at twice that channel spacing (122\,kHz, or 160\,m s$^{-1}$).  The remaining window used a coarse frequency resolution (128 channels of 15.628\,MHz width) to probe the 232\,GHz (1.3\,mm) continuum.  The observations alternated between AK Sco and the quasar J1709$-$3525 (3\degr\ separation) on a $\sim$7 minute cycle.  Some additional brief observations of the bright quasar J1626$-$2951 and Titan were made for calibration purposes.  The observing block lasted $\sim$1 hour, with half the time devoted to AK Sco.   

The raw visibility data were calibrated using the facility software package {\tt CASA} (v4.2).  After applying the standard system temperature and water vapor radiometer corrections, the intrinsic passband shape was determined using the observations of J1626$-$2951 and removed.  Complex gain variations due to instrumental and atmospheric effects were calibrated based on the regular monitoring of J1709$-$3525, and the overall amplitude scale was set using the observations of Titan.  After a single iteration of phase-only self-calibration, the reduced visibilities were time-averaged into 30\,s intervals.  The local continuum level was subtracted from the spectral windows containing emission lines of interest.  

Continuum and spectral line emission from all three of the targeted transitions were firmly detected.  The calibrated visibilities were Fourier inverted with natural weighting, deconvolved with the {\tt CLEAN} algorithm, and subsequently restored with a FWHM = $0\farcs75\times0\farcs60$ synthesized beam (with P.A. = 100\degr).  Some representative data products are shown together in Figure~\ref{fig:mom}.  

The synthesized map of the 1.3\,mm continuum emission (Fig.~1$a$) has an RMS noise level of 45\,$\mu$Jy beam$^{-1}$, and shows a marginally resolved morphology: an elliptical Gaussian fit to the visibilities indicates a total flux density of $32.65\pm0.07$\,mJy (a peak S/N $\approx 400$; an additional systematic uncertainty of $\sim$10\%\ is imposed by the absolute accuracy of the Titan emission model), with FWHM dimensions of $0\farcs38(\pm0.01)\times0\farcs12(\pm0.01)$ oriented at a position angle of $49\pm1\degr$.  The CO $J$=2$-$1 line is detected over a velocity span of $\sim$24\,km s$^{-1}$ (about 300 channels at the native resolution).  It spans a diameter of $\sim$2\arcsec, and has an integrated intensity of $2.21\pm0.01$\,Jy km s$^{-1}$.  For computational simplicity, we average these data into 60 channels of 305\,kHz (0.4\,km s$^{-1}$) width for further analysis.  The RMS noise level at that resolution is 4\,mJy beam$^{-1}$; the peak brightness is 170\,mJy beam$^{-1}$ (peak S/N $\approx 45$ per beam and channel).  With the same averaging, the $^{13}$CO and C$^{18}$O isotopologue transitions are also detected (albeit over a smaller velocity range and at lower significance).  Their integrated intensities are $0.54\pm0.01$ and $0.20\pm0.01$\,Jy km s$^{-1}$ (with peak S/N $\approx 12$ and 5 per beam and channel), respectively.

\section{CO Modeling and Results} \label{sec:method}

Our primary goal is to use the observed spectral visibilities that trace the CO\,$J$=2$-$1 emission line to quantitatively characterize the velocity field of the AK Sco gas disk, and thereby to measure the total mass of the central binary host.  We employed a forward-modeling approach that uses a parametric prescription for the disk structure (densities, temperatures, and dynamics), and then simulates the observed visibilities by assuming the molecular level populations are in local thermodynamic equilibrium (LTE) and propagating synthetic photons through the model structure with the {\tt RADMC-3D} radiative transfer code.\footnote{\url{ita.uni-heidelberg.de/~dullemond/software/radmc-3d/}}  Previous work with similar intentions \citep{simon00,pietu07,schaefer09,guilloteau14} has shown that resolved measurements of the disk rotation curve can constrain the central stellar mass ($M_{\ast}$) with high precision ($\sim$few percent, although with a linear dependence on distance).  
However, in practice this approach is complicated: it involves radiative transfer modeling with a large number of (unrelated) disk structure parameters.  Given that complexity, it is important to develop tests to validate its absolute accuracy.  \citet{rosenfeld12} used the circumbinary disk around the V4046 Sgr system to demonstrate that this approach is robust, in that it provides an $M_{\ast}$ estimate consistent with the constraints from spectroscopic monitoring of the stellar radial velocities.  The AK Sco system provides another rare opportunity to test the methodology.     

We adopted the two-dimensional (axisymmetric) parametric disk structure model that is described in detail by \citet{rosenfeld12}.  The temperature structure is vertically isothermal, and has a power-law radial distribution with index $q$ and a normalization at 10\,AU ($T_{10}$).  The radial surface density profile is the standard \citet{lynden-bell74} similarity solution for a viscous accretion disk, essentially a power-law with an exponential taper at large radii.  The gradient parameter $\gamma$ is fixed to unity, so the profile is described by a characteristic radius $r_c$ and a total CO mass $M_{\rm co}$.  The densities are distributed vertically under the assumption of hydrostatic equilibrium.  The bulk velocity field of the gas is assumed to be Keplerian, and dominated by the binary mass (assuming $M_{\rm disk}/M_{\ast} \ll 1$).  The line width is calculated as the quadrature sum of thermal and non-thermal (i.e., turbulent) broadening terms, with the latter denoted as a constant velocity width $\xi$.  The physical structure of the model is fully characterized by six free parameters, $\vt_{\rm disk} = \{T_{10}$, $q$, $M_{\rm co}$, $r_c$, $\xi$, $M_{\ast}$\}.

For a given set of these structure parameters, the {\tt RADMC-3D} radiative transfer code was used to calculate the corresponding molecular excitation levels \citep[appropriately assuming LTE for this transition; see][]{pavlyuchenkov07} and ray-trace synthetic model spectral images at high resolution (2.5\,AU pixels).  To do so, we specified an additional set of parameters.  The disk inclination $i_d$ is defined with respect to the rotation axis, such that $i_d = 0\degr$ corresponds to a face-on viewing geometry with the rotation axis pointing toward the observer (and the disk rotating counterclockwise), $i_d = 90\degr$ is an edge-on orientation, and $i_d = 180\degr$ is again face-on but with the rotation axis pointing away from the observer (and therefore an apparent clockwise sense of rotation).  The position angle $\varphi$ represents the projection of the angular momentum vector of the disk onto the sky (as typical, oriented E of N).  The disk center is characterized by offsets ($\delta_{\alpha}$, $\delta_{\delta}$) relative to the phase center.  Along with the distance $d$ and systemic velocity $v_{\rm sys}$, there are six additional free parameters, $\vt_{\rm obs} = \{i_d$, $\varphi$, $\delta_{\alpha}$, $\delta_{\delta}$, $v_{\rm sys}$, $d$\}.

The end result is a set of high resolution model images that specify the sky-projected CO\,$J$=2$-$1 surface brightness as a function of position and frequency for any set of parameters $\vt = \{\vt_{\rm disk}$, $\vt_{\rm obs}$\}.  We then employed the {\tt FFTW} algorithm \citep{fftw} to Fourier transform these spectral images, and then performed a band-limited interpolation of the complex visibilities onto the same spatial frequencies ($u$, $v$) sampled with ALMA \citep[using the spheroidal gridding functions advocated by][]{schwab84} to acquire a set of model visibilities $M_{u,v}(\vt)$.  To quantify the model quality given the data $D_{u,v}$, we computed a simple likelihood function
\begin{equation}
p(D | \vt) = \prod_{u,v} w_{u,v} \, |D_{u,v}-M_{u,v}(\vt)|^2, 
\end{equation}
where $w_{u,v}$ are the visibility weights (determined from the radiometer equation).  Eq.~(1) is identical to a log-likelihood function described by the sum of the standard $\chi^2$ values (real and imaginary) over all the observed spatial frequencies and velocity channels. 
To explore the posterior distribution of the model parameters, we employed the Markov Chain Monte Carlo (MCMC) technique with a Metropolis-Hastings (M-H) jump proposal.  

We assumed uniform priors in all parameters except $d$, since the ALMA data do not constrain the distance.  
A trigonometric parallax for AK Sco is available from {\it Hipparcos}, although the considerable optical variability is problematic: the original catalog has $d = 144^{+38}_{-25}$\,pc \citep{perryman97}, but the \citet{vanleeuwen07} revision argues for a closer value ($d = 102^{+26}_{-17}$\,pc).  \citet{pecaut12} measured a kinematic parallax corresponding to $d = 144\pm12$\,pc.  Recently, \citet{anthonioz15} combined astrometric and radial velocity monitoring to estimate $d = 141\pm7$\,pc.  We adopted a Gaussian prior on $d$ based on the weighted average of these latter two measurements, with mean 142\,pc and $\sigma = 6$\,pc.  

After a few preliminary chains were run, we empirically tuned the covariance of the M-H jumps to match the morphology of the posterior distribution, yielding a more efficient exploration of parameter space.  To determine the final posterior distributions, we ran multiple independent chains using different initializations.  After a conservative period of burn-in, we computed the Gelman-Rubin statistic \citep{gelman13} for each parameter over the ensemble of chains to assess convergence.\footnote{The code used to perform the analysis described here is open source and freely available under an MIT license at \url{https://github.com/iancze/JudithExcalibur}.}

\begin{figure*}[htb]
  \includegraphics[width=\textwidth]{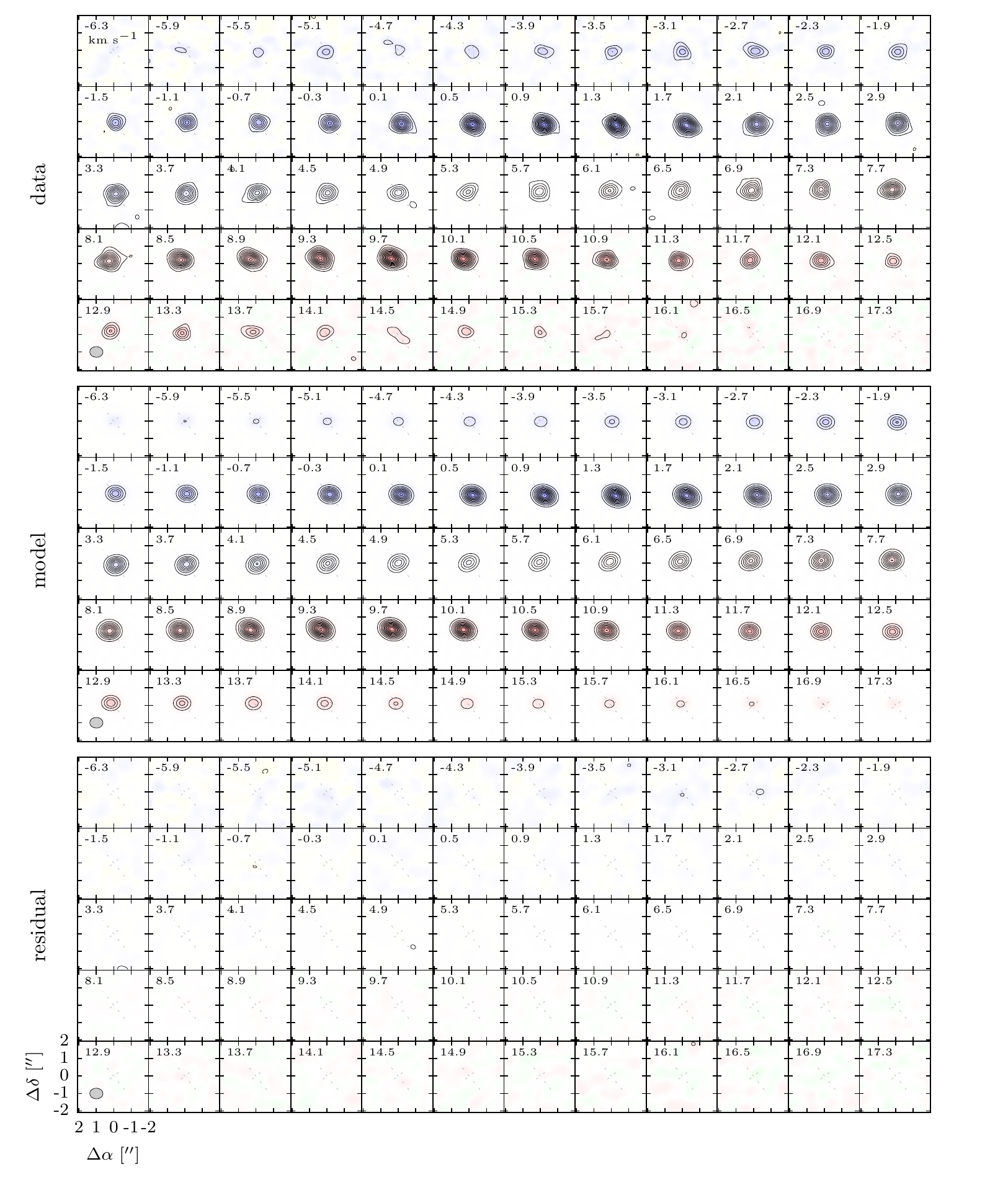}
  \figcaption{Channel maps at 305\,kHz (0.4\,km s$^{-1}$) resolution showing the CO $J$=2$-$1 data, the corresponding model, and the residuals (the latter two imaged in the same way as the data).  Contours are drawn at 3\,$\sigma$ (12\,mJy beam$^{-1}$) intervals, and the synthesized beam dimensions are shown in the lower left corners of each set of channel maps. The LSRK velocities are labeled in the top left of each panel. The color within each channel map corresponds to the velocity sampled in the moment maps (Figure~\ref{fig:mom}).
  \label{fig:results}}
\end{figure*}

\capstartfalse
\begin{deluxetable}{lr@{ $\pm$ }l|lr@{ $\pm$ }l}
  \tablecaption{\label{table:parameters}Inferred Parameters for AK Sco}
  \tablehead{\colhead{{\sc Parameter}} & \multicolumn{2}{c}{{\sc Value}} & \colhead{{\sc Parameter}} & \multicolumn{2}{c}{{\sc Value}}}
  \startdata
  $T_{10}$ (K) & 92   & 4    & $i_d$ (${}^\circ$)     & 109.4 & 0.5 \\
  $q$          & 0.51 & 0.01 & $\varphi$ (${}^\circ$) & 141.1 & 0.3 \\
  $\log M_\textrm{co}$ ($M_\oplus$) & -0.65 & 0.22  & $v_{\rm sys}$\tablenotemark{a} ($\kms$) & +5.49 & 0.06 \\
  $r_c$ (AU) & 14.3 & 1.2 & $\delta_\alpha$ (\arcsec) & 0.053  & 0.002 \\
  $M_\ast$ ($M_\odot$) & 2.49 & 0.10 & $\delta_\delta$ (\arcsec) & 0.045 & 0.002 \\
  $\xi$ ($\kms$) & 0.31 & 0.02 & $d$ (pc) & 143.6 & 5.7 
  \enddata
  \tablecomments{The quoted values represent the ``best-fit", the peaks of the marginal posterior  distributions.  The uncertainties correspond to the 68.3\%\ ($\sim$1\,$\sigma$) confidence intervals.}
  \tablenotetext{a}{In the LSRK frame, for the standard radio definition.  The corresponding heliocentric value is $-1.92\pm0.06$, consistent with the \citet{alencar03} value derived from optical spectroscopy.}
\end{deluxetable}
\capstarttrue

The parameter values inferred from modeling the observed spectral visibilities binned to 305\,kHz (0.4\,km s$^{-1}$) resolution are listed in Table 1.  A trial fit of the data at the best available (quasi-independent) spectral resolution of 122\,kHz (0.16\,km s$^{-1}$) recovered these same values.  The disk structure parameters are typical for similar Class II systems, although the characteristic radius and mass are on the small side.  Assuming a standard CO/H$_2$ abundance ratio ($\sim$10$^{-4}$), the total gas mass would be $\sim$0.007\,$M_{\odot}$ -- roughly consistent with what would be inferred from the continuum emission  \citep[e.g.,][]{andrews13}.  The inferred total mass of the central binary is $2.49\pm0.10$\,$M_{\odot}$, where the uncertainty is dominated by the distance prior.  At a fixed $d$, the constraint can be framed as $M_{\ast}/d = 0.01731\pm0.00015$\,$M_{\odot}$ pc$^{-1}$; i.e., the formal precision on the stellar mass is $\sim$1\%\ for a $\delta$-function prior on $d$.  Figure 2 shows a direct comparison of the data and best-fit model in the image plane, demonstrating (through the absence of significant residuals) the fit quality.  The reduced $\chi^2$ of the best-fit model is 1.08.      

\begin{figure}[htb]
  \includegraphics{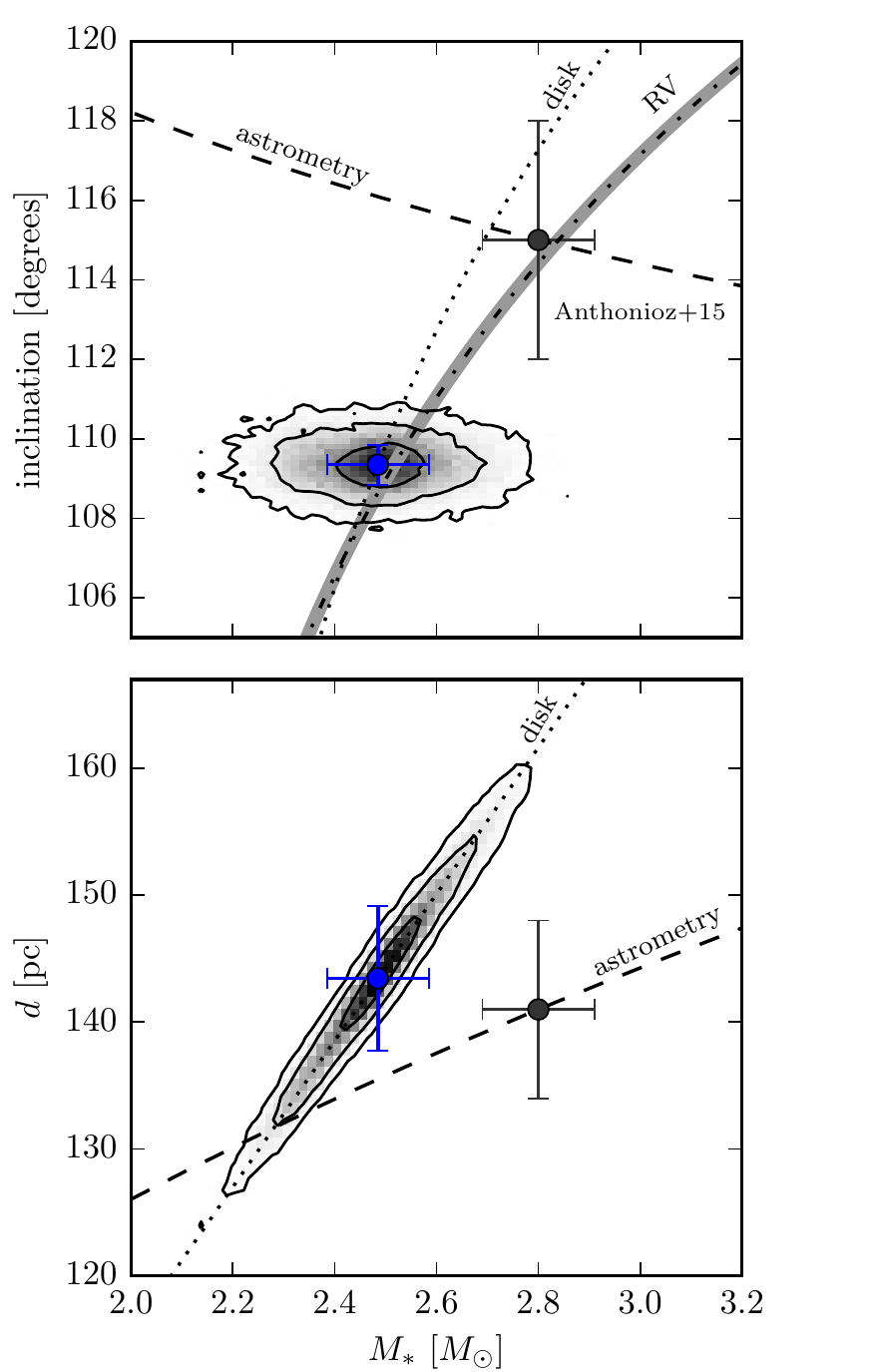}
  \figcaption{The (marginal) joint posterior probability distributions for $\{M_\ast$, inclination\} and $\{M_\ast, d\}$. For reference, we have overlaid the nominal parameter degeneracies ($\pm$1 $\sigma$) that accompany each measurement in isolation, assuming no external constraints about either parameter but precise knowledge of all other parameters. 
  \label{fig:triangle}}
\end{figure}

\section{Discussion}\label{sec:disc}

We have presented ALMA Cycle 1 observations around a wavelength of 1.3\,mm that resolve the dust and molecular line emission from the disk that orbits the double-lined spectroscopic binary AK Sco.  To our knowledge, these are the first observations that confirm the presence of a substantial molecular gas reservoir in this system.  The disk itself is relatively small (a characteristic radius of $\sim$14\,AU), but still contains a modest mass ($\sim$5--10\,$M_{\rm Jup}$ in total, for standard assumptions about opacities and abundances).  We have focused on a detailed modeling analysis of the spatially and spectrally resolved CO $J$=2$-$1 line emission, to map out the disk velocity field and make a dynamical estimate of the total mass of the host binary.  Adopting a well-motivated prior on the distance ($142\pm6$\,pc; see Sect.~3), we inferred a precise (4\%) combined stellar mass estimate given the ALMA spectral visibilities, $M_{\ast} = 2.49\pm0.10$\,$M_{\odot}$.

\subsection{Comparison to Previous Results}

There are two independent dynamical constraints on the AK Sco binary mass available in the literature.  The first comes from an optical spectroscopic campaign to monitor the radial velocity (RV) variations of the double-lined system, performed by \citet{alencar03}.  They measured a precise orbital period ($\sim$13.6\,d) and mass ratio ($0.987\pm0.005$), a substantial eccentricity ($e_{\ast} = 0.471\pm0.002$) and a total mass constraint $M_{\ast} \sin^3{i_{\ast}} = 2.114\pm0.010$\,$M_{\odot}$ where $i_{\ast}$ is the (unknown) inclination of the binary orbit.  These measurements do not depend on $d$.  If we assume the gas disk and binary orbits are co-planar ($i_{\ast} = i_d$), then our constraint on $i_d$ would convert the RV constraint on the binary mass to $2.52\pm0.03$\,$M_{\odot}$, identical (within 1\%) to our disk-based dynamical mass estimate.  These two dynamical mass constraints remain consistent (at the 1\,$\sigma$ level) so long as the binary and disk orbital planes are aligned within $\sim$2\degr, compatible with the statistical constraints on the mutual inclination distributions found for exoplanet systems \citep{tremaine12,figueira12,fabrycky14}. Additionally, by combining the distance-independent measurements of $M_{\ast}/d$ and $M_{\ast} \sin^3{i_{\ast}}$, we derive a dynamical distance to the system of $d=145.5 \pm 2.0$~pc, which is consistent with our initial assumptions in the form of the distance prior.

This consistency between mass constraints, along with the similar results for the V4046 Sgr binary \citep{rosenfeld12}, effectively validates the absolute accuracy of the disk-based dynamical mass measurement technique. Whereas the V4046 Sgr measurement involved a large circumstellar disk ($r_c = 45$ AU), we have demonstrated that---given the sensitivity and resolution of ALMA---the dynamical mass technique also works for much smaller disks like AK Sco ($r_c = 14$ AU). In the coming age of ALMA, this should be the workhorse approach for determining masses for large, statistically relevant, samples of young stars.  Most importantly, it is the {\it only} method capable of performing that task for {\it single} stars.

The second mass constraint comes from the recent work by \citet{anthonioz15}, which combined multi-epoch $H$-band interferometric observations with the \citet{alencar03} RV data.  In the context of a simple model that includes the binary and a narrow dust ring, these astrometric constraints suggest inclinations of $i_{\ast} = 115\pm3\degr$ and $i_\textrm{ring} = 121\pm8\degr$, respectively, which are $\sim$1.8 and 1.4\,$\sigma$ larger than the $i_d$ we infer from the ALMA data.  In conjunction with the RV constraints, this larger orbital inclination  suggests a higher binary mass, $M_{\ast} = 2.80\pm0.11$\,$M_{\odot}$, a 2\,$\sigma$ discrepancy with respect to the disk-based mass estimate.  For reference, Figure~\ref{fig:triangle} shows the mass constraints as a function of both inclination and distance.   

There are various scenarios that might alleviate this (modest) tension.  The difference between $i_\textrm{ring}$ and $i_d$ might be explained with a warp in the disk structure, which would perturb the line-of-sight projection of the velocity field measured in the CO line and thereby accommodate a higher $M_{\ast}$ estimate from the ALMA data \citep[e.g.,][]{rosenfeld12b,rosenfeld14,marino15}.  Given the eccentricity of the binary, one might also expect that the inclination estimates could be biased due to the assumptions of circular disk and ring structures.  For the \citeauthor{anthonioz15}~results, an imperfect ring model can affect the estimate of $i_{\ast}$ since the binary itself is not well-resolved (the semi-major axis of the apparent orbit is only 1 mas, about one third of the resolution).  We also explored an eccentric disk model, where the structure is built up from a series of infinitesimal apse-aligned elliptical rings: the ALMA data rule out a significant (mean) eccentricity in the gas disk, with $e_d < 0.04$ (at 99.7\%\ confidence).  In the end, it may just be that the (admittedly) simplistic ring model adopted in the preliminary analysis of \citet{anthonioz15} can be improved (the quoted reduced $\chi^2$ of their fit is $\sim$2), and such modifications would bring the results into agreement.  Ultimately, it would be interesting to combine all the measurements in a joint analysis.

\subsection{Comparison to Pre-MS Evolution Models}

To compare the disk-based dynamical mass with predictions from pre-MS evolution models, we performed the standard analysis of estimating stellar parameters from the Hertzsprung-Russell (H-R) diagram.  The $UBVRI$ photometry compilation of \citet{jensen97} was used to construct the AK Sco SED.  The near-infrared is contaminated by the dust disk, and so was excluded from our analysis.  A base parametric SED model was constructed from synthetic photometry in the {\sc BT-Settl} catalog of stellar models \citep{allard03}, interpolated for any given effective temperature ($T_{\rm eff}$) and surface gravity ($\log{g}$) assuming a fixed solar metallicity. This base model was adjusted for extinction using the \citet{fitzpatrick99} reddening curve with $R_V=4.3$ (as determined by \citet{manset05}) and then scaled by the squared ratio of the radius ($R_{\ast}$) and distance ($d$).  We assumed the components of the AK Sco binary are identical.  

\begin{figure}[ht!]
  \includegraphics{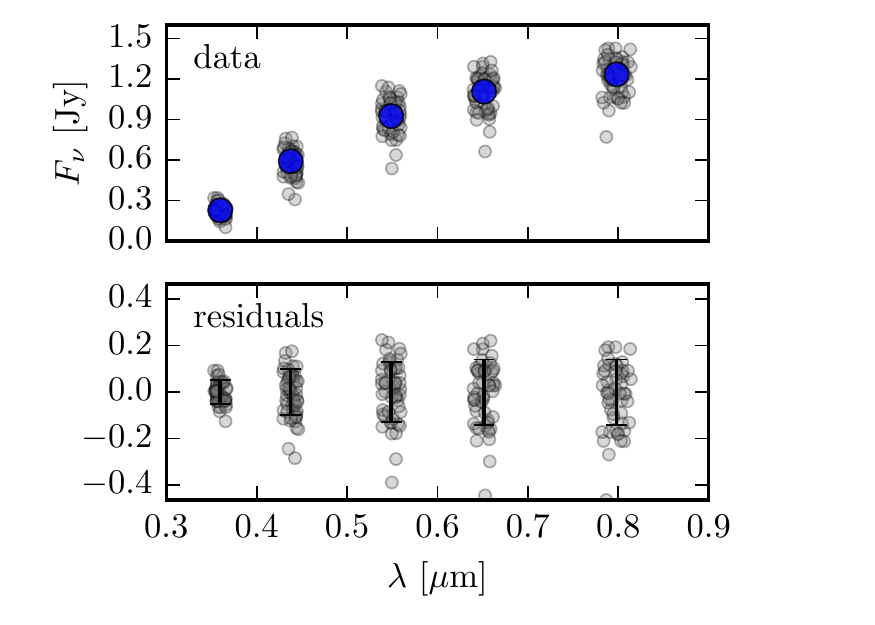}
  \includegraphics{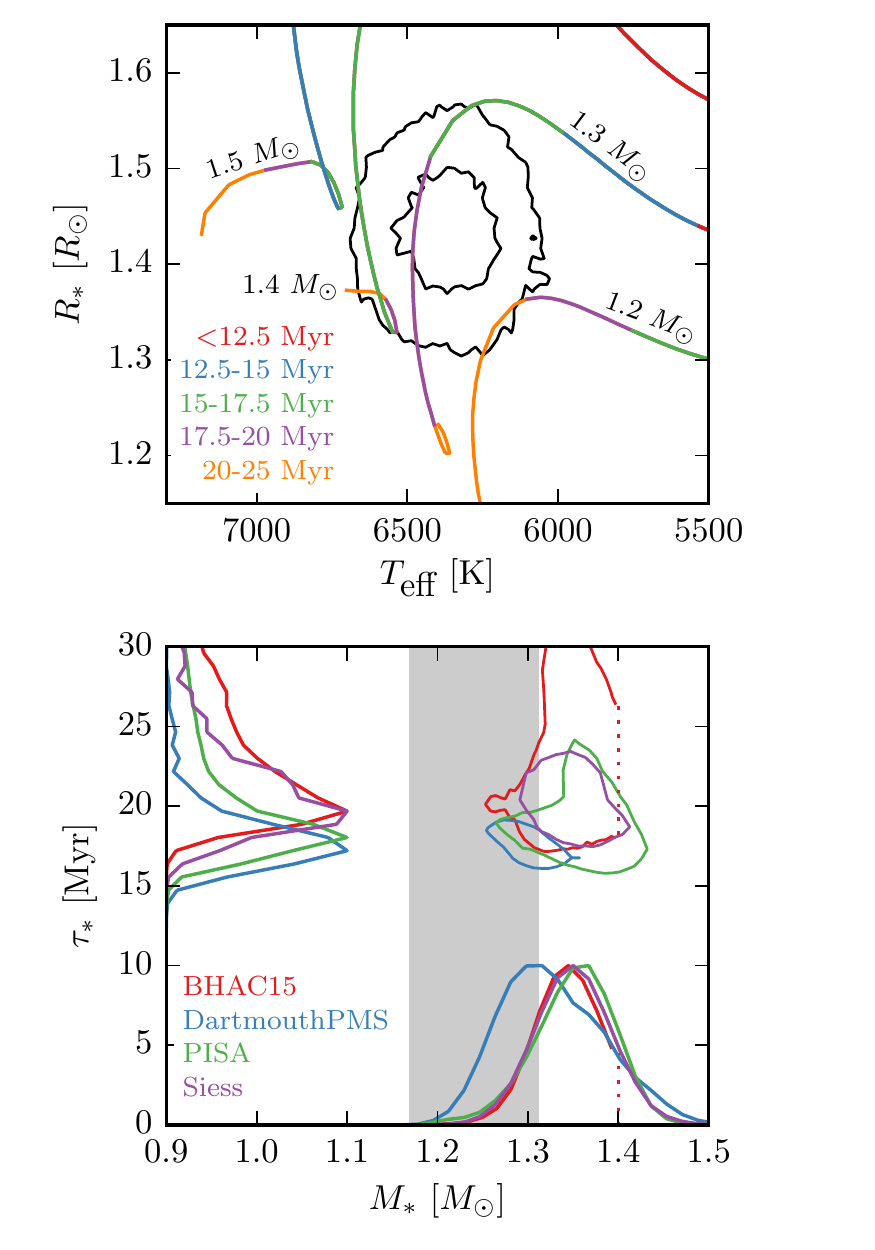}
  \figcaption{{\it top}: The SED compared with the best-fit model. The residuals are shown overlaid with the inferred ``jitter'' terms in each band to account for the systematic variance from variability. {\it middle}: The $R_{\ast}$, \teff\ version of the H-R diagram, with the marginal posteriors inferred from the SED modeling shown as 1 and 2\,$\sigma$ contours.  The \citet{dotter08} pre-MS model mass tracks are overlaid, with color scales indicating different ages (note that these values refer to each individual star in the AK Sco binary).  {\it bottom}: The joint mass and age constraints from various pre-MS models, shown as 1\,$\sigma$ contours and marginalized distributions.  The gray band marks the disk-based constraint on $M_{\ast}$ (1\,$\sigma$). The \citet{baraffe15} models do not explore the parameter space $M_\ast > 1.4 M_\odot$.
  \label{fig:PMS}}
\end{figure}

This SED model has five free parameters, $\vt_{\rm sed} = \{T_{\rm eff}$, $\log{g}$, $A_V$, $R_{\ast}$, $d$\}, but the photometric data alone cannot uniquely constrain all of them.  To aid in the inference of the stellar properties, we imposed simple Gaussian priors on $d$ (as in Sect.~3) and $T_{\rm eff}$.  The latter was based on the F5 spectral classification and de-reddened color indices \citep{andersen89}, which we associate with $T_{\rm eff} = 6450\pm150$\,K \citep[e.g.,][]{bessell79,popper80,gray92,casagrande10,pecaut13}.  AK Sco is highly variable, with erratic changes much larger than the uncertainties on individual measurements \citep[e.g.,][]{andersen89,alencar03}.  To deal with that variability, we incorporated a nuisance ``jitter" parameter ($\sigma$) at each band that characterizes the additional dispersion (assuming a Gaussian distribution).  With this parametric model setup, we explored the posterior distribution of $\vt = \{\vt_{\rm sed}$, $\sigma_U$, $\sigma_B$, $\sigma_V$, $\sigma_R$, $\sigma_I$\} conditioned on the SED data using MCMC with the ensemble sampler {\tt emcee} \citep{foreman-mackey13}.             

The modeling results are shown in Figure~\ref{fig:PMS}.  We found $T_{\rm eff} = 6365\pm155$\,K, $\log{g} = 3.5\pm0.5$, $R_{\ast} = 1.43\pm0.07$\,$R_{\odot}$, and $A_V = 0.70\pm0.1$\,mag for each star (and $d = 142\pm6$\,pc, as expected given the prior).  The corresponding luminosity of each component is $L_{\ast} = 3.0\pm0.5$\,$L_{\odot}$.  The variability dispersion terms, in terms of a flux density fraction, range from 0.1 ($I$-band) to 0.2 ($U$-band).     

Models of pre-MS evolution predict the joint behavior of \teff\ and $R_{\ast}$ (or equivalently $L_{\ast}$) as a function of $M_{\ast}$ and age ($\tau_{\ast}$).  Following the Bayesian formalism of \citet{jorgensen05}, we mapped the posterior constraints from the SED modeling into an inference on \{$M_{\ast}$, $\tau_{\ast}$\} for several evolutionary model grids \citep[for some practical examples, see][]{rosenfeld12,andrews13}.  This approach is also illustrated in Figure~\ref{fig:PMS}.  The AK Sco binary mass and age inferred from these model grids \citep{siess00,dotter08,tognelli11,baraffe15} are listed in Table 2.  Their weighted means and standard deviations are $M_{\ast} = 2.68\pm0.03$\,$M_{\odot}$ and $\tau_{\ast} = 18\pm1$\,Myr.  These model-predicted masses are consistent, but are all $\sim$1.5--2\,$\sigma$ higher than the dynamical mass inferred from the ALMA data.  The ages are also consistent with the age of the Upper Centaurus-Lupus association \citep[$\langle \tau_{\ast} \rangle = 16\pm1$\,Myr;][]{pecaut12}.         

\capstartfalse
\begin{deluxetable}{lr@{ $\pm$ }lr@{ $\pm$ }l}
  \tablecaption{\label{table:PMS}Evolutionary Model Predictions}
   \tablehead{\colhead{{\sc Model Grid}} & \multicolumn{2}{c}{$M_{\ast}$ [$M_\odot$]} & \multicolumn{2}{c}{$\tau_{\ast}$ [Myr]}}
  \startdata
  \citet{siess00} & 2.70 & 0.06 & 19 & 3\\
  \citet{tognelli11} & 2.71 & 0.07 & 18 & 2\\
  \citet{dotter08} & 2.60 & 0.07 & 17 & 2\\
  \citet{baraffe15} & 2.68 & 0.05 & 19 & 2
  \enddata
  \tablecomments{The quoted uncertainties correspond to the 68.3\%\ ($\sim$1\,$\sigma$) confidence intervals.}
\end{deluxetable}
\capstarttrue

The 7--10\%\ discrepancy between the pre-MS model masses and the disk-based dynamical mass is slightly larger than the typical level of disagreement noted for young EBs in this mass range \citep{Stassun2014}.  We suspect this modest mismatch might be attributed to the complexity of the AK Sco binary environment.  The AK Sco stars are separated by only $\sim$11\,$R_{\ast}$ at periastron (every $\sim$2 weeks), which leads to accretion bursts onto the stellar surface \citep{castro13}.  Perhaps related, the AK Sco stars have unusually broad ultraviolet lines that modulate with the binary period and are indicative of perturbed, hot ($\sim$60,000\,K) ``atmospheres" that extend out to 5\,$R_{\ast}$ \citep{castro09}.  This behavior is compounded by the erratic variability noted in broadband photometry and the Balmer emission lines \citep{alencar03}, suggesting additional complexity in the accretion/outflow behavior.  This dynamic, complicated environment likely impacts the physical structures and evolution of the stars, and is (obviously) not included in the pre-MS models. Alternatively, and perhaps more likely, the standard means of estimating parameters like \teff\ in such a situation should also be affected: a small shift ($<100$\,K cooler) would bring the measurements into agreement. 

\section{Summary and Conclusions} \label{sec:summary}

We have analyzed ALMA observations of the CO $J$=2$-$1 transition from the AK Sco circumbinary disk.  The main conclusions of this work include: \\

\noindent $\bullet$ A relatively compact disk in orbit around the AK Sco binary, with $\sim$5--10\,$M_{\rm Jup}$ of gas and dust, is detected in the 1.3\,mm continuum and main isotopologues of CO (the CO, $^{13}$CO, and C$^{18}$O $J$=2$-$1 transitions).  This suggests an unexpectedly long-lived ($\sim$18\,Myr) disk of primordial origin, as opposed to a second-generation debris disk. \\ 

\noindent $\bullet$ We employed a parametric disk structure model front-end fed into the radiative transfer code {\tt RADMC-3D} to generate synthetic spectral visibilities to compare with the observations in the MCMC framework.  The related software is provided as an open source resource to the community.  The results offer a high-quality dynamical measurement of the binary mass, $M_\ast = 2.49 \pm 0.10$\,$M_\odot$, that is independent of pre-MS evolution models. \\

\noindent $\bullet$ This disk-based dynamical mass estimate is in good agreement with the constraints from radial velocity monitoring of the binary, so long as the disk and binary orbital planes are aligned within $\sim$2\degr.  There is minor tension with a recent combined astrometric + radial velocity analysis, although we expect the comparison could be improved with a consistent joint analysis. \\

\noindent $\bullet$ With the standard approach of estimating stellar parameters from the H-R diagram, we make comparisons between pre-MS evolutionary model predictions and the dynamical mass estimated here.  These models suggest a slightly higher (by 7--10\%) stellar mass for AK Sco, in modest disagreement (at 1.5--2\,$\sigma$) with our results: this discrepancy may be attributed to the complicated accretion and interaction environment of the binary.  The model-dependent ages ($18\pm1$\,Myr) are consistent with the proposed AK Sco membership in the Upper Centaurus Lupus association. \\

\noindent $\bullet$ The overall consistency between $M_{\ast}$ estimates for AK Sco validates the absolute accuracy of the disk-based dynamical mass technique.  This method has great promise in the ALMA era, since it is uniquely capable of providing precise (few \%) masses of statistically large samples of {\it single} pre-MS stars that can be used to test and calibrate models of early stellar evolution.

\acknowledgments
The authors appreciate the helpful advice of Eric Stempels and Maxwell Moe; help with an eccentric disk model from Meredith Hughes, Sam Factor, and Eugene Chiang; and the technical support regarding our many ALMA calibration questions from Jennifer Donovan Meyer. IC is supported by the NSF Graduate Fellowship and the Smithsonian Institution.  SA acknowledges the very helpful support provided by the NRAO Student Observing Support program related to the early development of this project.  This paper makes use of the following ALMA data: ADS/JAO.ALMA\#2012.1.00496.S.  ALMA is a partnership of ESO (representing its member states), NSF (USA), and NINS (Japan), together with NRC (Canada) and NSC and ASIAA (Taiwan), in cooperation with the Republic of Chile.  The Joint ALMA Observatory is operated by ESO, AUI/NRAO, and NAOJ.  Figure~\ref{fig:triangle} was generated with \texttt{triangle.py} \citep{foreman-mackey14}. This research made extensive use of the
Julia programming language \citep{julia12} and Astropy \citep{astropy13}.

\end{document}